\documentclass[twocolumn,showpacs,preprintnumbers,amsmath,amssymb,superscriptaddress]{revtex4}

\usepackage{graphicx}

\usepackage{dcolumn}

\usepackage{bm}

\begin{document}


	\title{The QED Radiative Corrections to Chiral Magnetic Effect}

	\author{Bo  Feng}
	\affiliation{School of Physics, Huazhong University of Science and Technology, Wuhan 430074, China}
	
	\author{De-fu Hou}
	\affiliation{Institute of Particle Physics, Huazhong Normal University, Wuhan 430079, China}
	
	\author{Hai-cang Ren}
	\affiliation{	Physics Department, The Rockefeller University, 1230 York Avenue, New York, New York 10021-6399, USA}
	\affiliation{Institute of Particle Physics, Huazhong Normal University, Wuhan 430079, China}
	
	\date{\today}


	\begin{abstract}
		We study in this paper the radiative corrections to chiral magnetic current at both zero and nonzero temperature. Our motivation is a radiative correction to the matrix element 
		of the anomalous Ward identity in massless QED stemming from a three-loop diagram where the two photons coming from the one-loop anomalous triangle are re-scattered.  
		Through the interplay between the Ward identity and the infrared subtlety of the fermion loop integral, we are able to reproduce the corrections known in literature in a simpler approach and obtain its contribution to the chiral magnetic current at zero temperature. At a nonzero temperature, the infrared subtlety disappeared in a static magnetic field and the three-loop diagram does not contribute to the chiral magnetic current any more. The generalization to all orders of the massless QED and the QCD corrections are discussed. 
	\end{abstract}





	\maketitle

	\section{Introduction}

	Anomaly induced transport phenomena in systems with chiral fermions have attracted wide interests ranging from high energy physics to condensed matter physics\cite{Reviews}.
	One simple anomalous transport phenomenon induced by chiral anomaly is the chiral magnetic effect(CME)\cite{CME1,CME2,CME3}, which predicts charge asymmetries in the final stage of the relativistic heavy ion collisions\cite{CMEinHIC1,CMEinHIC2,CMEinHIC3,CMEinHIC4,CMEinHIC5} and negative magnetoresistance in some Weyl semimetals\cite{Weylsemi1,Weylsemi2}. While there are evidences of CME in condensed matter physics, on the heavy ion collision side, it remains to exclude the noisy backgrounds in order to nail down the real CME signal\cite{taskforce1,taskforce2,taskforce3}.

	The chirality imbalance necessary for implementing CME in high energy physics is provided by the axial charge fluctuations, which is induced by the tunneling among 
	topologically inequivalent gluon configurations of Quantum Chromodynamics(QCD) at high temperature. In spite of the difficulty in a systematic study of the creation process of the chirality imbalance in the early stage 
	of heavy ion collisions and the subsequent generation of CME in a transient magnetic field, one can model the chirality imbalance by a  nonzero axial chemical potential $\mu_5$, which corresponds to the temporal component of an external axial vector field. To the first order in an external electromagnetic potential $A_\mu$ and a constant $\mu_5$, the chiral magnetic current can be written as\cite{CME3,RenandHou}
	\begin{equation}
		J_i(q)=\eta\mu_5K_{ij}(q)A_j(q),
		\label{kernel}
	\end{equation}
	where the factor $\eta=1$ for electrons and $\eta=3\sum_fq_f$ for quark-gluon plasma(QGP) with $q_f$ the electric charge quanta of the quark flavor $f$. 
	In terms of the AVV three-point function, $\Lambda_{\mu\nu\rho}(Q_1,Q_2)$,  standing for the proper vertex of two external photons and an external axial vector field  with outgoing photon momenta $Q_1$ and $Q_2$,  the kernel $K_{ij}(q)=\Lambda_{ij4}(q,-q)$  corresponding to a constant temporal component of the external axial vector field. The lowest order of $\Lambda_{\mu\nu\rho}(Q_1,Q_2)$, denoted by $\Delta_{\mu\nu\rho}(Q_1,Q_2)$, is represented by the one-loop triangle diagrams. Therefore, it is tempting to relate the CME kernel $K_{ij}(q)$ to the axial anomaly 
	via the limiting process
	\begin{equation}
		K_{ij}({q})=-i\lim_{k_0\rightarrow 0}\frac{1}{k_0}(Q_1+Q_2)_\rho\Delta_{ij\rho}(Q_1,Q_2).\label{tentative}
	\end{equation} 
	with $Q_1=({\bf q}, i(\omega+k_0/2)), Q_2=(-{\bf q}, i(-\omega+k_0/2))$ and $\Delta_{\mu\nu\rho}$ the amplitudes of triangle diagrams. From the well-known anomalous Ward identities\cite{Adler,AdlerandBardeen},  
	\begin{equation}
		(Q_1+Q_2)_\rho\Delta_{\mu\nu\rho}(Q_1,Q_2)=-i\frac{e^2}{2\pi^2}\epsilon_{\mu\nu\alpha\beta}Q_{1\alpha}Q_{2\beta},
		\label{anomaly}
	\end{equation}
	it follows that
	\begin{equation}
		K_{ij}(q)=i\frac{e^2}{2\pi^2}\epsilon_{ikj}q_k,
	\end{equation}
	and the classical form of the chiral magnetic current
	\begin{equation}
		{\bf J}=\frac{e^2}{2\pi^2}\mu_5{\bf B}.\label{classicalCME}
	\end{equation} 
	emerges for one flavor and one color degrees of freedom in a static and uniform magnetic field. Because of the non-renormalization theorem of chiral anomalies, it is 
	expected that eq.(\ref{classicalCME}) is free from radiative corrections.

	Twenty years after the discovery of the non-renormalization theorem of chiral anomalies, the authors of \cite{Ansel'm} found a radiative correction to (\ref{anomaly}) for massless QED, 
	which is UV divergent. This correction is contributed by the three-loop diagram in Fig.1, where the two photons from the upper triangle are rescattered and the anomalous Ward identity 
	is then modified to the form
	\begin{align}
		\nonumber (Q_1+Q_2)_\rho\Lambda_{\mu\nu\rho}(Q_1,Q_2)
		=&-i\frac{e^2}{2\pi^2}\epsilon_{\mu\nu\alpha\beta}Q_{1\alpha}Q_{2\beta}\\
		&\times\left(1-\frac{3 e^4}{64\pi^4}\log\frac{\Lambda^2}{k^2}\right),
		\label{radiativecorrection}
	\end{align} 
	with $\Lambda$ the ultraviolet cut-off and $k$ the infrared cutoff depending on $Q_1$ and $Q_2$. This observation is non-trivial in heavy ion collisions, wherein the light $u$ and $d$ quarks can be approximately regarded as massless and (\ref{radiativecorrection}) generates a radiative correction to the expected chiral magnetic current via
	\begin{equation}
		K_{ij}(q)=i\frac{e^2}{2\pi^2}\mu_5\epsilon_{ikj}q_k\left(1-\frac{3 e^4}{64\pi^4}\log\frac{\Lambda^2}{q^2}\right).
	\end{equation}
	where the UV cutoff may be related to an energy scale when the chemical potential description of the axial charge imbalance breaks down.
	In this paper, we shall calculate the contribution from the lowest order photon-photon scattering diagrams as in Fig. 1 to the chiral magnetic current in massless fermion case at zero temperature and a nonzero temperature. At zero temperature, we are able to 
	re-produce the $e^6$ terms on RHS of (\ref{radiativecorrection}) following a simpler approach. Our method is based on the interplay between the vector Ward identities and the infrared
	behavior of the diagram and can be applied to the nonzero temperature case as well.

	On the other hand, the constant $\mu_5$ limit in (\ref{tentative}) becomes subtle at a nonzero temperature. For more general $Q_1$ and $Q_2$, say $Q_1=({\bf q}+{\bf k}/2,i(\omega+k_0/2))$
	and $Q_1=(-{\bf q}+{\bf k}/2,i(-\omega+k_0/2))$ with $k=({\bf k},ik_0)$ the 4-momentum carried by $\mu_5$, the limits ${\bf k}\to 0$ and $k_0\to 0$ do not commute as a symptom of the 
	Lorentz symmetry breaking. To one-loop order, we have\cite{CME3,RenandHou}
	\begin{equation}
		\lim_{k_0\to 0}\lim_{{\bf k}\to 0}\Lambda_{ij4}(Q_1,Q_2)=i\frac{e^2}{2\pi^2}\mu_5\epsilon_{ikj}q_k\left[1+{\cal O}\left(\frac{{\bf q}^2}{T^2}\right)\right],
		\label{order1}
	\end{equation}
	but
	\begin{equation} 
		\lim_{{\bf k}\to 0}\lim_{k_0\to 0}\Lambda_{ij4}(Q_1,Q_2)=i\frac{e^2}{2\pi^2}\mu_5\epsilon_{ikj}q_k\times{\cal O}\left(\frac{{\bf q}^2}{T^2}\right).
		\label{order2}
	\end{equation}

	To assess the contribution of the three-loop diagram Fig.1 at a finite temperature, both orders of limits have to be considered. In order to take the zero energy 
	limit, one has to start with a real time Green's function, whose infrared behavior is more convoluted to track. Fortunately, after taking the limit (\ref{order1}) or (\ref{order2}) of 
	the top triangle in Fig.1, the real time formulation can be converted into Matsubara formulation for a static magnetic field, where the infrared behavior is transparent, 
	and the same form of the limits (\ref{order1}) or (\ref{order2}) for the summation of the one-loop and three-loop diagrams emerges. In another word, the term on RHS of (\ref{order1}) that is linear in ${\bf q}$ 
	is not subject to the radiative corrections of Fig.1 and the classical formula (\ref{classicalCME}) is intact.

	The rest of the paper is organized as follows: in Section II we calculate the radiative correction from Figure.1 at zero temperature. In section III, we discuss the contributions from Figure.1  at finite temperature. Section IV is devoted to the conclusion along with some open issues. Some technical calculation details and useful proof are presented in the appendices. Throughout the paper, we will work with Euclidean signature and a 4-momentum is represented by ${q_\mu=({\bf q}, iq_0)}$ with $q_0$ the real energy. All gamma matrices in this paper are 
	hermitian.




	\begin{figure}
		\begin{center}
			\includegraphics[height=6cm]{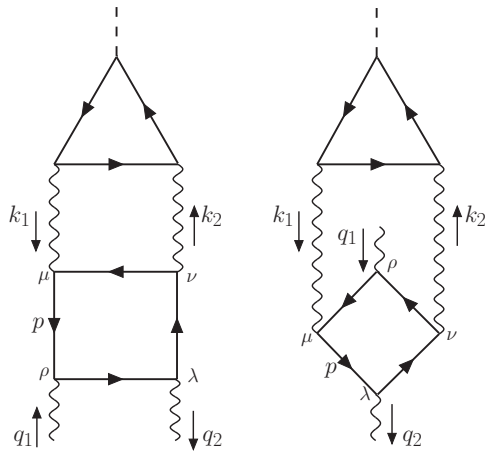}
		\end{center}
		\caption{\label{fig:epsart} The photon-photon scattering contributions to chiral anomaly. The dashed and wavy lines represent the axial-vector and vector fields, respectively.}
	\end{figure}

	\section{Zero-temperature Analysis}

	To assess higher order contributions, it is convenient to distinguish the anomalous Ward identity at the operator level and its matrix elements. In case of electron-photon 
	or quark-photon system, the diagram of the AVV proper vertex can be decomposed into a ladder of two-photon irreducible parts linked by two photon lines shown in Fig.2, where a photon line stands for a dressed photon propagator. The Anomalous Ward identity at the operator level reads
	\begin{equation}
		\partial^\mu j_\mu^5=2imj^5+i\frac{\alpha_0}{4\pi}\epsilon_{\rho\sigma\lambda\nu}F_{\rho\sigma}F_{\lambda\nu}.\label{triang}
	\end{equation}
	with $\alpha_0$ the bare coupling constant, $F_{\rho\sigma}$ the electromagnetic field strength tensor, $j_\mu^5=i\bar\psi\gamma_\mu\gamma_5\psi$ and $j^5={\bar\psi}\gamma_5\psi$. The operator anomaly corresponds to the 4-divergence with respect to the axial vector vertex of the shaded triangle at the leftmost in Fig.2, whose anomaly term on RHS is free from radiative corrections as proved by Adler and Bardeen\cite{AdlerandBardeen}. In another word, besides the one-loop triangle, the fermion loop with the axial vector vertex of all other diagrams included in the shaded triangle have more than two photon vertices 
	so the shift of loop momentum does not generate extra terms. The entire diagram is obtained by taking the matrix element of eq.(\ref{triang}) between the vacuum and a state with two photons of momenta ($Q_1$, $Q_2$) and polarizations $(\rho, \lambda)$. The matrix element takes the form
	\begin{align}
		\nonumber &(Q_1+Q_2)_\mu\Lambda_{\mu\rho\lambda}(Q_1,Q_2)\\
		\nonumber =&-i\left[2mG\left(\frac{Q_1^2}{m^2},\frac{Q_2^2}{m^2},\frac{Q_1\cdot Q_2}{m^2}\right)+H\left(\frac{Q_1^2}{m^2},\frac{Q_2^2}{m^2},\frac{Q_1\cdot Q_2}{m^2}\right)\right]\\
		&\times\epsilon_{\rho\lambda\alpha\beta}Q_{1\alpha}Q_{2\beta},
		\label{adlerbardeen}
	\end{align}   
	with $G$ corresponding to the naive divergence and $H$ to the anomaly coefficient. The Adler-Bardeen theorem\cite{AdlerandBardeen} states that the naive divergence $2mj^5$ has a known value
	to all orders at a particular kinematic point where external momenta vanish with a nonzero mass, i.e.
	\begin{equation}
		2mG(0,0,0)+H(0,0,0)=0, \qquad H(0,0,0)=\frac{2\alpha}{\pi}.
		\label{adlerbardeen_}
	\end{equation}
	with $\alpha$ the renormalized coupling constant. The first equation above follows from the Sutherland-Veltman theorem\cite{SutherlandVeltman} and the second one states that the photon-photon scattering contribution in Fig.1 vanishes at this kinematic point.  
	On the contrary,  however, the kinematic point that validate the Adler-Bardeen theorem cannot be attained if zero fermion mass limit is taken first, which renders the arguments of 
	$G$ and $H$ approaching infinity.  Therefore, the discovery that the matrix element of the anomaly, $H$, acquires a radiative correction is not surprising, nor does it 
	invalidate the Adler-Bardeen theorem\cite{Adlerreview}.

	Both Sutherland-Veltman theorem and Adler-Bardeen theory follow from the Ward identity of the electromagnetic gauge invariance. In the rest of this section, we shall show that the 
	Ward identity argument failed in the presence of the infrared divergence pertaining to massless fermions. Our analysis also gives rise to a much simpler derivation of the radiative 
	correction in (\ref{radiativecorrection}). A more pedagogical evaluation of the diagrams in Fig. 1 can be found in Appendix A. To make the infrared singularities transparent, we shall work with imaginary energy formulation for the rest of this section. This amounts to a Wick rotation of all 4-momenta (external and internal). The real energy result can be obtained by undoing the Wick rotation of the external momenta appropriately.

	It follows from (\ref{anomaly}) that the divergence with respect to the axial vector vertex of the top triangle in Fig.1 can be replaced by a two photon vertex specified by the 
	RHS of (\ref{anomaly}). The three-loop contribution to the 4-divergence $(Q_1+Q_2)_\mu\Lambda_{\mu\rho\lambda}(Q_1,Q_2)$ of (\ref{adlerbardeen}) is thereby reduced to a two-loop 
	diagram whose amplitude reads
	\begin{equation}
		\Lambda_{\rho\lambda}(q,Q)=i\frac{e^6}{2\pi^2}\epsilon_{\mu\nu\alpha\beta}Q_\alpha\int\frac{d^4k}{(2\pi)^4}\frac{k_\beta}{(k^2)^2}\Gamma_{\mu\nu\rho\lambda}(q,q;k),\label{3loopCME}
	\end{equation}
	to the leading order in $Q$, where $Q_1=-q+Q/2$, $Q_2=q+Q/2$ and $\Gamma_{\mu\nu\rho\lambda}(q_1,q_2;k_1)$ is the amplitude of the box diagram of four photons in Fig.3. We have
	\begin{equation}
		\Gamma_{\mu\nu\rho\lambda}(q_1,q_2;k_1)=\rm{I}+\rm{II}+\rm{III}+\rm{IV}+\rm{V}+\rm{VI},
	\end{equation} 
	with 
	\begin{align}
		\rm{I} =& -\int\frac{d^4p}{(2\pi)^4}{\rm tr}\gamma_\mu\frac{1}{p\!\!\!/-{k\!\!\!/}_1}\gamma_\nu\frac{1}{p\!\!\!/-{q\!\!\!/}_2+{q\!\!\!/}_1}\gamma_\rho
		\frac{1}{p\!\!\!/-{q\!\!\!/}_2}\gamma_\lambda\frac{1}{p\!\!\!/},\nonumber
	\end{align}
		\begin{align}
		\rm{II} =& -\int\frac{d^4p}{(2\pi)^4}{\rm tr}\gamma_\mu\frac{1}{p\!\!\!/-{k\!\!\!/}_1}\gamma_\lambda\frac{1}{p\!\!\!/-{k\!\!\!/}_1+{q\!\!\!/}_2}\gamma_\nu
		\frac{1}{p\!\!\!/+{q\!\!\!/}_1}\gamma_\rho\frac{1}{p\!\!\!/},\nonumber\\
		\rm{III} =& -\int\frac{d^4p}{(2\pi)^4}{\rm tr}\gamma_\mu\frac{1}{p\!\!\!/}\gamma_\rho\frac{1}{p\!\!\!/-{q\!\!\!/}_1}\gamma_\lambda\frac{1}{p\!\!\!/-{q\!\!\!/}_1+{q\!\!\!/}_2}\gamma_\nu\frac{1}{p\!\!\!/+{k\!\!\!/}_1},\nonumber\\
		\rm{IV} =& -\int\frac{d^4p}{(2\pi)^4}{\rm tr}\gamma_\mu\frac{1}{p\!\!\!/-{k\!\!\!/}_1}\gamma_\nu\frac{1}{p\!\!\!/-{q\!\!\!/}_2+{q\!\!\!/}_1}\gamma_\lambda
		\frac{1}{p\!\!\!/+{q\!\!\!/}_1}\gamma_\rho\frac{1}{p\!\!\!/},\nonumber\\
		\rm{V} =& -\int\frac{d^4p}{(2\pi)^4}{\rm tr}\gamma_\mu\frac{1}{p\!\!\!/}\gamma_\rho\frac{1}{p\!\!\!/-{q\!\!\!/}_1}\gamma_\nu\frac{1}{p\!\!\!/+{k\!\!\!/}_1-{q\!\!\!/}_2}\gamma_\lambda\frac{1}{p\!\!\!/+{k\!\!\!/}_1},\nonumber\\
		\rm{VI} =& -\int\frac{d^4p}{(2\pi)^4}{\rm tr}\gamma_\mu\frac{1}{p\!\!\!/}\gamma_\lambda\frac{1}{p\!\!\!/+{q\!\!\!/}_2}\gamma_\rho\frac{1}{p\!\!\!/-{q\!\!\!/}_1+{q\!\!\!/}_2}\gamma_\nu\frac{1}{p\!\!\!/+{k\!\!\!/}_1}.\label{amplitude}
	\end{align}

	It is well know that $\Gamma_{\mu\nu\rho\lambda}(q_1,q_2;k_1)$ is UV convergent because of the gauge invariance. A regulator is yet required to maintain the gauge invariance in general since an individual terms of (\ref{amplitude}) remain logarithmically divergent. 
	To see this, let us contract a photon vertex with the 4-momentum of the photon, say, $q_{1\rho}\Gamma_{\mu\nu\rho\lambda}(q_1,q_2;k_1)$. 
	Following the standard textbook approach \cite{Peskin}, the integrand of $q_{1\rho}\Gamma_{\mu\nu\rho\lambda}(q_1,q_2;k_1)$ can be written as 
	the difference of two terms, differing by the shift of the integration momentum. Each term itself is linearly divergent by power counting and regulator is
	required to render the shift of the integration momentum legitimate such that the Ward identity holds. On the other hand, Upon contracting with the epsilon tensor from the anomaly, 
	each term of difference in the integrand of $\epsilon_{\mu\nu\alpha\beta}q_{1\rho}\Gamma_{\mu\nu\rho\lambda}(q_1,q_2;k_1)$ becomes logarithmic divergent and the regulator is no longer required to maintain the Ward identity(see Appendix C for details). Also,
	the difference by shifting integration variable by an external momentum in the integrand of $q_{1\rho}\Gamma_{\mu\nu\rho\lambda}(q_1,q_2;k_1)$ can be Taylor expanded in the powers of the external momenta, leaving the integrand a total derivative with respect to the integration momentum.

	It follows that
	\begin{align}
		{\rm I}+{\rm II}+{\rm IV} =& i\int\frac{d^4p}{(2\pi)^4}\frac{\partial}{\partial p_\lambda}{\rm tr}\gamma_\mu\frac{1}{p\!\!\!/-{k\!\!\!/}_1}
		\gamma_\nu\frac{1}{p\!\!\!/+{q\!\!\!/}_1}\gamma_\rho\frac{1}{p\!\!\!/},\nonumber\\
		{\rm III}+{\rm V}+{\rm VI} =& i\int\frac{d^4p}{(2\pi)^4}\frac{\partial}{\partial p_\lambda}{\rm tr}\gamma_\mu\frac{1}{p\!\!\!/}
		\gamma_\rho\frac{1}{p\!\!\!/-{q\!\!\!/}_1}\gamma_\nu\frac{1}{p\!\!\!/+{k\!\!\!/}_1},
		\label{group1}
	\end{align}
	at $q_2=0$ and 
	\begin{align}
		\nonumber	&{\rm I}+{\rm IV}+{\rm V} \\
		\nonumber	=& i\int\frac{d^4p}{(2\pi)^4}\frac{\partial}{\partial p_\rho}{\rm tr}\gamma_\mu\frac{1}{p\!\!\!/-{k\!\!\!/}_1}
		\gamma_\nu\frac{1}{p\!\!\!/-{q\!\!\!/}_2}\gamma_\lambda\frac{1}{p\!\!\!/}+\Xi_{\mu\nu\rho\lambda},\\
		\nonumber	&{\rm II}+{\rm III}+{\rm VI}\\
		=& i\int\frac{d^4p}{(2\pi)^4}\frac{\partial}{\partial p_\rho}{\rm tr}\gamma_\mu\frac{1}{p\!\!\!/}
		\gamma_\lambda\frac{1}{p\!\!\!/+{q\!\!\!/}_2}\gamma_\nu\frac{1}{p\!\!\!/+{k\!\!\!/}_1}+\Xi_{\mu\nu\rho\lambda}^\prime.
		\label{group2}
	\end{align}
	at $q_1=0$, where
	\begin{widetext}	
		\begin{align}
	   	\Xi_{\mu\nu\rho\lambda}=\int\frac{d^4p}{(2\pi)^4}{\rm tr}\gamma_\mu\left(
			\frac{1}{p\!\!\!/-{k\!\!\!/}_1}\gamma_\rho\frac{1}{p\!\!\!/-{k\!\!\!/}_1}\gamma_\nu\frac{1}{p\!\!\!/-{q\!\!\!/}_2}\gamma_\lambda\frac{1}{p\!\!\!/}
			-\frac{1}{p\!\!\!/}\gamma_\rho\frac{1}{p\!\!\!/}\gamma_\nu\frac{1}{p\!\!\!/+{k\!\!\!/}_1-{q\!\!\!/}_2}\gamma_\lambda\frac{1}{p\!\!\!/+{k\!\!\!/}_1}
			\right)=0,
		\end{align}
		and
		\begin{align}
			\Xi_{\mu\nu\rho\lambda}^\prime=\int\frac{d^4p}{(2\pi)^4}{\rm tr}\gamma_\mu \left(
			\frac{1}{p\!\!\!/}\gamma_\lambda\frac{1}{p\!\!\!/+{q\!\!\!/}_2}\gamma_\nu\frac{1}{p\!\!\!/+{k\!\!\!/}_1}\gamma_\rho\frac{1}{p\!\!\!/+{k\!\!\!/}_1}-\frac{1}{p\!\!\!/-{k\!\!\!/}_1}\gamma_\lambda\frac{1}{p\!\!\!/-{k\!\!\!/}_1+{q\!\!\!/}_2}\gamma_\nu\frac{1}{p\!\!\!/}\gamma_\rho\frac{1}{p\!\!\!/}
			\right)=0.
		\end{align}
		upon shifting the integration momentum and we end up with the total derivative forms of the integrands of $\Gamma_{\mu\nu\rho\lambda}(q_1,0;k_1)$ and 
		$\Gamma_{\mu\nu\rho\lambda}(0,q_2;k_1)$. In particular, 
		\begin{align}
			\nonumber		\epsilon_{\mu\nu\alpha\beta}\Gamma_{\mu\nu\rho\lambda}(0,0;k_1) =& i\int\frac{d^4p}{(2\pi)^4}\frac{\partial}{\partial p_\lambda}\epsilon_{\mu\nu\alpha\beta}{\rm tr}\gamma_\mu
			\left(\frac{1}{p\!\!\!/-{k\!\!\!/}_1}\gamma_\nu\frac{1}{p\!\!\!/}\gamma_\rho\frac{1}{p\!\!\!/}
			+\frac{1}{p\!\!\!/}\gamma_\rho\frac{1}{p\!\!\!/}\gamma_\nu\frac{1}{p\!\!\!/+{k\!\!\!/}_1}\right),\\
			=& i\int\frac{d^4p}{(2\pi)^4}\frac{\partial}{\partial p_\rho}\epsilon_{\mu\nu\alpha\beta}{\rm tr}\gamma_\mu
			\left(\frac{1}{p\!\!\!/-{k\!\!\!/}_1}\gamma_\nu\frac{1}{p\!\!\!/}\gamma_\lambda\frac{1}{p\!\!\!/}
			+\frac{1}{p\!\!\!/}\gamma_\lambda\frac{1}{p\!\!\!/}\gamma_\nu\frac{1}{p\!\!\!/+{k\!\!\!/}_1}\right).
			\label{leading} 
		\end{align}
		While it is tentative to employ the Stokes' theorem to convert the volume integral into a surface integral at infinity of the 
		momentum space, care must be exercised because of the UV and IR behavior of the integrand. Introducing a sphere of radius $\epsilon$ around each point of IR singularity, 
		i.e. $p=0$ and $p=\pm k_1$ and applying the Stokes theorem outside the spheres. The expressions under each partial derivative vanishes faster than 
		$p^{-3}$ as $p\to\infty$. So the surface integral at infinity does not contribute and we are left with the surface integral of each sphere as well as the volume integral 
		over the interior of them. The integrand of the surface integral grows like $p^{-2}$ or $(p\mp k_1)^{-1}$ while the corresponding volume integral grows like 
		$p^{-3}$ or $(p\mp k_1)^{-2}$. Consequently these integrals vanish as $\epsilon\to 0$ and we obtain that   
		\begin{equation}
			\epsilon_{\mu\nu\alpha\beta}\Gamma_{\mu\nu\rho\lambda}(0,0,k)=0.\label{epsilonbox}
		\end{equation}
		Taking the derivatives of (\ref{group1}) with respect to $q_1$, and (\ref{group2}) with respect to $q_2$, we obtain that
		\begin{equation}
			\left.\frac{\partial}{\partial q_{1\sigma}}\Gamma_{\mu\nu\rho\lambda}(q_1,0;k)\right|_{q_1=0}
			=\int\frac{d^4p}{(2\pi)^4}\frac{\partial}{\partial p_\lambda}{\rm tr}\gamma_\mu\left(
			-\frac{1}{p\!\!\!/-k\!\!\!/}\gamma_\nu\frac{1}{p\!\!\!/}\gamma_\sigma\frac{1}{p\!\!\!/}\gamma_\rho\frac{1}{p\!\!\!/}
			+\frac{1}{p\!\!\!/}\gamma_\rho\frac{1}{p\!\!\!/}\gamma_\sigma\frac{1}{p\!\!\!/}\gamma_\nu\frac{1}{p\!\!\!/+k\!\!\!/}
			\right),
			\label{gamma_1}
		\end{equation}
		and 
		\begin{align}
			\left.\frac{\partial}{\partial q_{2\sigma}}\Gamma_{\mu\nu\rho\lambda}(0,q_2;k)\right|_{q_2=0}
			=\int\frac{d^4p}{(2\pi)^4}\frac{\partial}{\partial p_\rho}{\rm tr}\gamma_\mu\left(
			\frac{1}{p\!\!\!/-k\!\!\!/}\gamma_\nu\frac{1}{p\!\!\!/}\gamma_\sigma\frac{1}{p\!\!\!/}\gamma_\lambda\frac{1}{p\!\!\!/}
			-\frac{1}{p\!\!\!/}\gamma_\lambda\frac{1}{p\!\!\!/}\gamma_\sigma\frac{1}{p\!\!\!/}\gamma_\nu\frac{1}{p\!\!\!/+k\!\!\!/}
			\right).
			\label{gamma_2}
		\end{align}

		It follows from (\ref{gamma_1}) and (\ref{gamma_2}) that the coefficient of the linear term in $q$ of $\Gamma_{\mu\nu\rho\lambda}(q,q,k)$ reads
		\begin{align}
			\left.\frac{\partial}{\partial q_\sigma}\Gamma_{\mu\nu\rho\lambda}(q,q;k)\right|_{q=0}
			=&\left.\frac{\partial}{\partial q_{1\sigma}}\Gamma_{\mu\nu\rho\lambda}(q_1,0;k)\right|_{q_1=0}+\left.\frac{\partial}{\partial q_{2\sigma}}\Gamma_{\mu\nu\rho\lambda}(0,q_2;k)\right|_{q_2=0},\nonumber\\
			=&-\int\frac{d^4p}{(2\pi)^4}\left[\frac{\partial}{\partial p_\lambda}{\rm tr}\gamma_\mu\left(
			\frac{1}{p\!\!\!/-{k}\!\!\!/}\gamma_\nu\frac{1}{p\!\!\!/}\gamma_\sigma\frac{1}{p\!\!\!/}\gamma_\rho\frac{1}{p\!\!\!/}
			-\frac{1}{p\!\!\!/}\gamma_\rho\frac{1}{p\!\!\!/}\gamma_\sigma\frac{1}{p\!\!\!/}\gamma_\nu\frac{1}{p\!\!\!/+{k}\!\!\!/}
			\right)-(\rho\leftrightarrow\lambda)\right]. \label{integrand}
		\end{align}
		Let us focus on the integration of the first term inside the parenthesis and applying the Stokes theorem outside the small spheres centered at $p=0$ and $p=k$ as before.
		The surface integral at infinity drops and the singularity at $p=k$ is too weak to contribute. The integration around the singularity $p=0$, however, has to be retained.
		Denoting the small sphere centered at the origin by $B_\epsilon$ and its surface by $\partial B_\epsilon$, we end up with
		\begin{align}
			\nonumber 	J_{\mu\nu\rho\lambda\sigma} \equiv & -\int\frac{d^4p}{(2\pi)^4}\frac{\partial}{\partial p_\lambda}{\rm tr}\gamma_\mu\frac{1}{p\!\!\!/-{k}\!\!\!/}
			\gamma_\nu\frac{1}{p\!\!\!/}\gamma_\sigma\frac{1}{p\!\!\!/}\gamma_\rho\frac{1}{p\!\!\!/},
			\\
			=& \frac{1}{(2\pi)^4}\int_{\partial B_\epsilon} d^3\hat p_\lambda\frac{{\rm tr}\gamma_\mu(p\!\!\!/-{k}\!\!\!/)\gamma_\nu{\hat p}\!\!\!/\gamma_\sigma{\hat p}\!\!\!/
				\gamma_\rho{\hat p}\!\!\!/}{(p-k)^2}
			-\int_{B_\epsilon}\frac{d^4p}{(2\pi)^4}\frac{\partial}{\partial p_\lambda}{\rm tr}\gamma_\mu\frac{1}{p\!\!\!/-{k}\!\!\!/}
			\gamma_\nu\frac{1}{p\!\!\!/}\gamma_\sigma\frac{1}{p\!\!\!/}\gamma_\rho\frac{1}{p\!\!\!/},\label{J}
		\end{align}
		where $\hat p$ denotes the unit vector in the direction of $p$. As $\epsilon\to 0$, $p-k\to-k$ and
		\begin{align}
			\epsilon_{\mu\nu\alpha\beta}Q_\alpha k_\beta J_{\mu\nu\rho\lambda\sigma} 
			=& -\frac{1}{8\pi^2k^2}\epsilon_{\mu\nu\alpha\beta}Q_\alpha k_\beta\langle{\hat p}_\lambda{\rm tr}\gamma_\mu k\!\!\!/\gamma_\nu{\hat p}\!\!\!/
			\gamma_\sigma{\hat p}\!\!\!/\gamma_\rho{\hat p}\!\!\!/\rangle_p\nonumber\\
			&+ \frac{1}{8\pi^2k^2}\epsilon_{\mu\nu\alpha\beta}Q_\alpha k_\beta\int_0^\epsilon dpp^3\left\langle\frac{\partial}{\partial p_\lambda}
			\frac{{\rm tr}\gamma_\mu k\!\!\!/\gamma_\nu p\!\!\!/\gamma_\sigma p\!\!\!/ \gamma_\rho p\!\!\!/}{(p^2)^3}\right\rangle_p.
			\label{integral}
		\end{align}
	\end{widetext}
	where $\langle...\rangle_p$ denotes the average over the direction of $p$. We have
	\begin{align}
		&\left\langle{\hat p}_\mu{\hat p}_\nu\right\rangle_p = \frac{1}{4}\delta_{\mu\nu},\nonumber\\
		&	\left\langle{\hat p}_\mu{\hat p}_\nu{\hat p}_\rho{\hat p}_\lambda\right\rangle_p = \frac{1}{24}
		(\delta_{\mu\nu}\delta_{\rho\lambda}+\delta_{\mu\rho}\delta_{\nu\lambda}+\delta_{\mu\lambda}\delta_{\nu\rho}).
		\label{symmetry} 
	\end{align}
	It is readily seen that the second integral of (\ref{integral}) vanishes because
	\begin{align}
		\left\langle\frac{\partial}{\partial p_\lambda}\frac{p_\mu p_\nu p_\rho}{(p^2)^3}\right\rangle_p=0,
	\end{align}
	and we are left with
	\begin{align}
		\nonumber	&\epsilon_{\mu\nu\alpha\beta}Q_\alpha k_\beta J_{\mu\nu\rho\lambda\sigma} \\
		=& -\frac{1}{8\pi^2k^2}\epsilon_{\mu\nu\alpha\beta}Q_\alpha k_\beta\left\langle{\hat p}_\lambda{\rm tr}\gamma_\mu k\!\!\!/\gamma_\nu{\hat p}\!\!\!/
		\gamma_\sigma{\hat p}\!\!\!/\gamma_\rho{\hat p}\!\!\!/\right\rangle_p,\nonumber\\
		\nonumber	=& \frac{1}{96\pi^2k^2}\epsilon_{\mu\nu\alpha\beta}Q_\alpha k_\beta k_\tau \\
		&\times{\rm tr}\gamma_\mu\gamma_\tau\gamma_\nu
		(\gamma_\lambda\gamma_\sigma\gamma_\rho+\gamma_\rho\gamma_\lambda\gamma_\sigma+\gamma_\sigma\gamma_\rho\gamma_\lambda).
	\end{align}
	where the second step follows from (\ref{symmetry}) and the following two identities of gamma matrices
	\begin{align}
		&\gamma_\alpha\gamma_\mu\gamma_\alpha = -2\gamma_\mu,\nonumber\\
		&	\gamma_\alpha\gamma_\mu\gamma_\nu\gamma_\rho\gamma_\alpha = -2\gamma_\rho\gamma_\nu\gamma_\mu,
	\end{align}
	Further reduction is facilitated by the identity
	\begin{equation}
		\epsilon_{\mu\nu\alpha\beta}\gamma_\mu\gamma_\tau\gamma_\nu=2(\delta_{\tau\alpha}\gamma_\beta-\delta_{\tau\beta}\gamma_\alpha)\gamma_5,
	\end{equation}
	and we obtain that
	\begin{align}
		\nonumber &\epsilon_{\mu\nu\alpha\beta}Q_\alpha k_\beta J_{\mu\nu\rho\lambda\sigma}\\
		=&\frac{1}{16\pi^2k^2}(k\cdot Qk_\alpha-k^2 Q_\alpha){\rm tr}\gamma_5\gamma_\sigma\gamma_\rho\gamma_\lambda\gamma_\alpha,\nonumber\\
		=&\frac{1}{4\pi^2k^2}\epsilon_{\sigma\rho\lambda\alpha}(k\cdot Qk_\alpha-k^2 Q_\alpha).\label{epsilonJ}
	\end{align}





	
It is straightforward to show that all four terms of the integrand of (\ref{integrand}) contribute equally, i.e.
\begin{equation}
	\left.\frac{\partial}{\partial q_\sigma}\Gamma_{\mu\nu\rho\lambda}(q,q;k)\right|_{q=0}=4J_{\mu\nu\rho\lambda\sigma},
	\label{coeff4}
\end{equation}	 
and the Taylor expansion of 
$\epsilon_{\mu\nu\alpha\beta}Q_\alpha k_\beta\Gamma_{\mu\nu\rho\lambda}(q,q;k)$ to the linear order in $q$ reads
\begin{align}
	\epsilon_{\mu\nu\alpha\beta}Q_\alpha k_\beta\Gamma_{\mu\nu\rho\lambda}(q,q;k) 
	=\frac{1}{\pi^2k^2}q_\sigma\epsilon_{\sigma\rho\lambda\alpha}(k\cdot Qk_\alpha-k^2 Q_\alpha).
	\label{box}
\end{align}
following (\ref{epsilonbox}), (\ref{coeff4}) and (\ref{epsilonJ}). 
	Substituting (\ref{box}) into (\ref{3loopCME}), the integration over $k$ diverges logarithmically in both UV and IR regions. The former is genuine and its 
	coefficient matches that of $\ln\Lambda^2$ in (\ref{radiativecorrection}). The infrared divergence, however, is caused by the inappropriate small $q$ expansion as 
	$k\to 0$. It follows from (\ref{amplitude}) that $\Gamma_{\mu\nu\rho\lambda}(q,q;k)$ tends to a finite limit as $k\to 0$ with $q$ in the denominator of the integrand, 
	which serves an infrared cutoff. The radiative corrections in (\ref{radiativecorrection}) is thereby reproduced without tedious multiloop calculations. The finite term 
	pertaining to the logarithm cannot be obtained that simply.   As discussed in\cite{Adler,AdlerandBardeen}, this divergence can be removed through the renormalization of the axial-vector vertex.

	Note that the same diagrams as that in Figure.1 with the two internal photon lines replaced by gluon lines may contribute to the chiral magnetic current  more significantly because of the strong coupling strength. The kernel of the chiral magnetic current in this case becomes \footnote{In this case, $e^4$ coming from the QED vertices pertaining to the internal photon lines is replaced by $\frac{1}{2N_c}(N_c^2-1)g_{\rm YM}^4$ for the color group $SU(N_c)$ with $g_{\rm YM}$ the Yang-Mills coupling.}
	\begin{align}
		K_{ij}(q)=i\frac{e^2}{2\pi^2}\mu_5\epsilon_{ikj}q_k\left(1-\frac{g^4}{32\pi^4}\log\frac{\Lambda^2}{q^2}\right).
	\end{align} 
	with $g$ the QCD coupling constant.

	\begin{figure}
		\begin{center}
			\includegraphics[height=7.0cm]{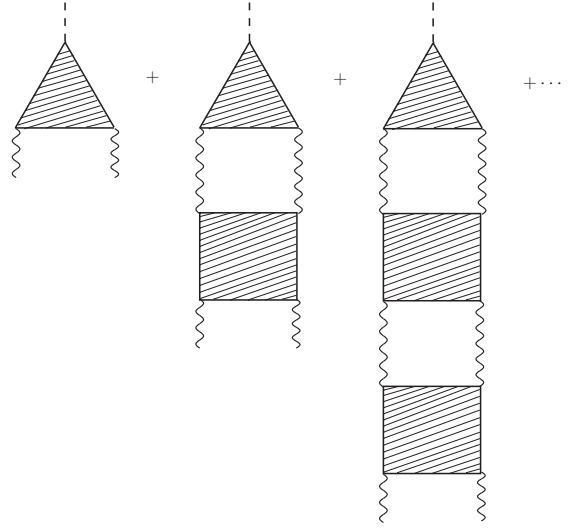}
		\end{center}
	\caption{\label{fig:epsart} The two-photon \textit{reducible} AVV diagrams. The shaded box and triangle represent the two-photon \textit{irreducible} diagrams.}
	\end{figure}

	\begin{figure}
\begin{center}
			\includegraphics[height=6.0cm]{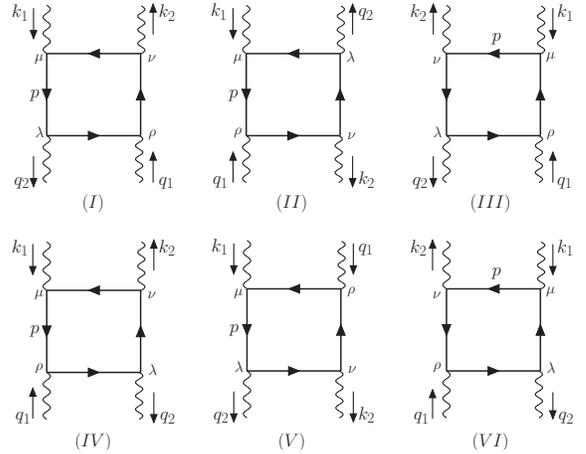}
		\end{center}
		\caption{\label{fig:epsart} The four-photon box diagrams including all permutations.}
	\end{figure}

	\section{Finite-temperature Analysis}

	In this section, we shall discuss whether the same radiative corrections to CME current from the two-photon reducible diagrams remains at a nonzero temperature. The problem is subtle because of the Lorentz symmetry breaking at a nonzero temperature, which may cause the limit $q\to 0$ of the 4-momentum $q$ flowing in a vertex ambiguous. That is, $\lim_{q_0\rightarrow 0}\lim_{|{\bf q}|\rightarrow 0}$ and $\lim_{|{\bf q}|\rightarrow 0}\lim_{q_0\rightarrow 0}$, may not agree. The ambiguity already shows up at one-loop calculations of CME current with $q$ the 4-momentum entering the vertex of the axial chemical potential or the vertex attached to the external magnetic field. We shall examine the two orders of limit separately for the former case in what follows. To carry out the zero energy limit involved, one has to work with a real time formulation, of which the closed-time-path (CTP) Green's functions\cite{CTP} serves our purpose. 
	As in the case of zero temperature, we designate $q_1$ to the incoming 4-momentum of the external vector potential and $q_2$ to the outgoing momentum of the electric current. 
	The incoming 4-momentum of the axial chemical potential is then $Q=q_2-q_1$.

	\subsection{The CTP Green's function}

	The CTP Green's functions are generated by a path integral whose action is the integration of the classical Lagrangian along a closed time path which consists of a forward branch, $\int_{-\infty}^\infty dt(...)$ and a backward branch, 
	$\int_{\infty}^{-\infty} dt(...)$. The number of degrees of freedom is thereby doubled. Explicitly the path integral that generates the CTP Green's function is given by
	\begin{align}
		\nonumber Z_{\rm CTP}=&\int[d\phi_1(t)][d\phi_2(t)]e^{iS_{\rm CTP}[\phi_1(t),\phi_2(t)]}\\
		&\times W[\phi_2(-\infty),\phi_1(-\infty)],
	\end{align}
	with the action 
	\begin{align}
		\nonumber &S_{\rm CTP}[\phi_1(t), \phi_2(t)]\\
		=&\int_{-\infty}^\infty dtL_{\rm cl.}[\phi_1(t)]+\int_\infty^{-\infty}dtL_{\rm cl.}[\phi_2(t)],
		\label{ctpaction}
	\end{align}  
	where $\phi$ stands for the set of all field variables and 
	\begin{equation}
		W[\phi',\phi]=<\phi'|\rho|\phi>.
	\end{equation}
	generates initial correlations with $\rho$ a density operator. As the path integral for the ordinary Feynman diagrams, the interaction is adiabatically switched on since $t=-\infty$.    
	While the CTP Green's function can be described by the same set of Feynman diagrams, the number of components of a $n$-point CTP Green's function is $2^n$ times of a Green's function of Feynman because of the additional CTP indices pertaining to external lines. It follows from (\ref{ctpaction}), a bare CTP vertex is formed by the field variables along the same time branch and carries one CTP index, while an internal line, a CTP propagator, carries two CTP indices and has four components. All CTP indices not attached to external lines have to be 
	summed.  Given all eight CTP components of the diagram for the chiral magnetic effect, the electric current in response to an external magnetic field and an axial chemical potential is obtained by fixing the CTP index of the current vertex on the forward time 
	branch and summing up all CTP indices of the magnetic field and axial chemical potential vertices, which take the same values along 
	both time branches. A photon vertex in CTP formulation reads $\gamma_\mu$ ($-\gamma_\mu$) on the forward (backward) time branch with the minus sign taking into account the opposite direction of the time integration in (\ref{ctpaction}). Likewise for the axial vector vertex, i.e., $\gamma_\mu\gamma_5$ and $-\gamma_\mu\gamma_5$. With an equilibrium density matrix at $t=-\infty$, 
	$\rho\propto e^{-\beta H_0}$ with $H_0$ the free Hamiltonian, the four CTP components $S_{ab}(p|m)$ of a free Diract propagator of mass $m$ 
	takes the form
	\begin{align}
		S_{11}(p|m) =& \frac{i}{p\!\!\!/+i0^+-m}-\pi\frac{p\!\!\!/+m}{E}\nonumber\\ &
		\times\left[f(E)\delta(p_0-E)+f(E)\delta(p_0+E)\right],\nonumber\\
		S_{12}(p|m) =& -\pi\frac{p\!\!\!/+m}{E}\lbrace f(E)\delta(p_0-E)\nonumber\\ &
		+[f(E)-1]\delta(p_0+E)\rbrace,\nonumber\\
		S_{21}(p|m) =& -\pi\frac{p\!\!\!/+m}{E}[f(E)-1]\delta(p_0-E)\nonumber\\ &
		+f(E)\delta(p_0+E)\rbrace,\nonumber\\
		S_{22}(p|m) =& \frac{-i}{p\!\!\!/-i0^+-m}-\pi\frac{p\!\!\!/+m}{E}
		[f(E)\delta(p_0-E)\nonumber\\ &+f(E)\delta(p_0+E)].
		\label{ctpS}
	\end{align}
	where $p=(\mathbf{p},ip_0)$, $E=\sqrt{{\mathbf p}+m^2}$, $p\!\!\!/\equiv -i\gamma_\nu p_\nu$ and
	$f(x)=1/(e^{\beta x}+1)$ is the Fermi distribution function at temperature $T=1/\beta$.
	Let $\Lambda_{\mu\nu\rho}^{abc}(Q_1,Q_2)$ be a CTP component of the three point function corresponding to $\Lambda_{\mu\nu\rho}(Q_1,Q_2)$ at zero temperature with the index pair $(\mu,a)$ associated to the electric current, the index pair $(\nu,b)$ to the external electromagnetic potential and the index pair $(\rho, c)$ to the axial vector vertex. The retarded Green's function underlying the current in response to an external electromagnetic potential and an external axial vector field reads
	\begin{equation}
		\Lambda_{\mu\nu\rho}^R(Q_1,Q_2)=\sum_{b,c}\Lambda_{\mu\nu\rho}^{1bc}(Q_1,Q_2).
		\label{retarded}
	\end{equation}
	We have $\mu=i,\nu=j$, $\rho=4$, $Q_1=-q_1$ and $Q_2=q_2$ for the radiative correction to CME by the diagrams in Fig.1.

	After working out the double limit $\lim_{Q_0\to 0}\lim_{\bf Q\to 0}$ or $\lim_{\bf Q\to 0}\lim_{Q_0\to 0}$, the rest of the calculation can be greatly simplified for the standard setup 
	of chiral magnetic current, i.e. the spatial current under a static and homogeneous magnetic field. The summation (\ref{retarded}) yields the Matsubara amplitude, 
	i.e. the different CTP components in (\ref{retarded}) add up to a single term which is described by the same diagram with all continuous energies flowing along internal lines 	replaced by discrete Matsubara energies. The loopwise proof of this statement is rather tedious. A nonperturbative proof in terms of spectral representations is presented 	in Appendix B.

	\subsection{The order $\lim_{Q_0\to 0}\lim_{\bf Q\to 0}$:}

	The first limit, ${\bf Q}\to 0$ renders the axial chemical potential homogeneous and we have ${\bf q}_1={\bf q}_2\equiv {\bf q}$. The CTP formulation has to be employed to evaluate 
	the second limit,  $Q_0\to 0$.
	Let us start with an excursion to prove that the divergence of the AVV triangle in CTP formulation can be replaced by a bare CTP vertex, like in the zero temperature case.
	To evaluate the fermion loop, it is convenience to introduces the matrix notation with CTP indices traced together with the spinor and internal indices.
	Introducing the $2\times 2$ matrix with respect to the CTP indices, i.e.,
	\begin{equation}
		\eta_1=\left(\begin{array}{cc} 1 & 0\\
			0 & 0\\ \end{array}\right), 
		\ \ \ \ \ \hbox{and} 
		\qquad \eta_2=\left(\begin{array}{cc} 0 & 0\\
			0 & 1\\ \end{array}\right).
		\label{projection}
	\end{equation}
	different CTP components can be projected out with one of them. It can be shown explicitly that(see appendix D for a general proof without relying on the thermal equilibrium)
	\begin{align}
		\nonumber S(p+q|m)Q\!\!\!/\gamma_5\eta_cS(p|m)=& i[\eta_c\gamma_5S(p|m)+S(p+q|m)\gamma_5\eta_c]\\
		&+2mS(p+q|m)\Gamma_5\eta_cS(p|m),
		\label{basic}
	\end{align}
	with
	\begin{equation} 
		Q\!\!\!/\equiv -iq_\mu\Gamma_\mu .
	\end{equation}
	An arbitrary CTP component of the Pauli-Villars regularized one loop AVV triangle reads
	\begin{widetext}
		\begin{align}
			\nonumber 		&\Delta_{\mu\nu\rho}^{abc}(Q_1,Q_2)\\
			=&-\int\frac{d^4p}{(2\pi)^4}{\rm Tr}\Gamma_\rho\gamma_5\eta_c\left[S(p-Q_2|0)\Gamma_\nu\eta_bS(p|0)\Gamma_\mu\eta_aS(p+Q_1|0)
			+S(p-Q_1|0)\Gamma_\mu\eta_aS(p|0)\Gamma_\nu\eta_bS(p+Q_2|0)\nonumber\right.\\
			&-\left.S(p-Q_2|M)\Gamma_\nu\eta_bS(p|M)\Gamma_\mu\eta_aS(p+Q_1|M)-S(p-Q_1|M)\Gamma_\mu\eta_aS(p|M)\Gamma_\nu\eta_bS(p+Q_2|M)\right],
		\end{align}
		with the trace extending to both spinor and CTP indices. Taking the divergence with respect to the axial vector vertex, we have
		\begin{align}
			\nonumber		&i(Q_1+Q_2)_\rho\Delta_{\rho\mu\nu}^{cab}(Q_1,Q_2)\\
			=&-i\int\frac{d^4p}{(2\pi)^4}\left[F_{\mu\nu}^{cab}(p-Q_2,p)-F_{\mu\nu}^{cab}(p,p+Q_2)
			+F_{\nu\mu}^{cba}(p-Q_1,p)-F_{\nu\mu}^{cba}(p,p+Q_1)\right]\nonumber\\
			&+2M\int\frac{d^4p}{(2\pi)^4}{\rm Tr}\Gamma_5\eta_c\left[S(p-Q_2|M)\Gamma_\nu\eta_bS(p|M)\Gamma_\mu\eta_aS(p+Q_1|M)
			+S(p-Q_1|M)\Gamma_\mu\eta_aS(p|M)\Gamma_\nu\eta_bS(p+Q_2|M)\right],
			\label{divergence}
		\end{align}
		with $F_{\mu\nu}^{cab}(p,q)\equiv{\rm Tr}\left[\Gamma_\mu\gamma_5\eta_c\eta_aS(p|0)\Gamma_\nu\eta_bS(q|0)-\Gamma_\mu\gamma_5\eta_c\eta_aS(p|M)\Gamma_\nu\eta_bS(q|M)\right]$, 
		where the identity (\ref{basic}) has been employed.
		The second and fourth terms of the integrand in the first line of (\ref{divergence}) differ from the first and third terms by shifting the integration 
		momentum and UV behavior of each term is regularized a la Pauli-Villars. The integration thereby vanish and we end up with the contribution from the regulator only, i.e.
		\begin{align}
			\nonumber 	i(Q_1+Q_2)_\rho\Delta_{\mu\nu\rho}^{abc}(Q_1,Q_2)=&2M\int\frac{d^4p}{(2\pi)^4}{\rm Tr}\Gamma_5\eta_c\left[S(p-Q_2|M)\Gamma_\nu\eta_bS(p|M)\Gamma_\mu\eta_aS(p+Q_1|M)\right.
			\\
			&+\left.S(p-Q_1|M)\Gamma_\mu\eta_aS(p|M)\Gamma_\nu\eta_bS(p+Q_2|M)\right].
		\end{align}
		In the limit $M\to\infty$, all Fermi distribution functions embedded in CTP propagators can be dropped. Among the eight CTP components (111, 112, 121, 211, 122,
		212, 221, 222), the 111-component correspond to the Feynman diagram at zero temperature and generates the standard Adler-Bardeen anomaly. Because of the anti-time 
		ordering (opposite sense of the Wick rotation), the contribution of the 222-component takes the negative value of the 111-component. As to other six CTP components, 
		they all carry the combination of the form  $S_{12}(p|M)S_{21}(p+q|M)$ with $q$ the external momentum entering the vertex $S_{12}$ and $S_{21}$.  It follows from (\ref{ctpS})
		that in the limit $M\to\infty$ the energy integration of these components 
		\begin{equation}
			\int_\infty^\infty dp_0\delta(p_0+E_{\bf p})\delta(p_0+q_0-E_{{\bf p}+{\bf q}})(...)=\delta(E_{\bf p}+E_{\bf p+\bf q}-q_0)(...)\to 0.
		\end{equation}
		Consequently, only the 111 and 222 CTP components of the AVV triangle divergence left over and can be coded in a two-point CTP vertex 




		%



		\begin{equation}
			(Q_1+Q_2)_\rho\Delta_{\mu\nu\rho}^{abc}(Q_1,Q_2) =
			\left\{
			\begin{aligned}
				& -i\frac{e^2}{4\pi^2}\epsilon_{\mu\nu\rho\lambda}Q_{1\rho}Q_{2\lambda},\  a=b=c=1 \, ; \\
				& i\frac{e^2}{4\pi^2}\epsilon_{\mu\nu\rho\lambda}Q_{1\rho}Q_{2\lambda},\  a=b=c=2 \, ; \\
				& 0, \hbox{otherwise} .
			\end{aligned}
			\right.
			\label{anomaly_}
		\end{equation}


		Inserting (\ref{anomaly_}) with $Q_1=k_1$, $Q_2=-k_2$ and ${\bf k}_1={\bf k}_2$ into the 4-divergence of the CTP amplitude of the diagram Fig.1, we obtain, to the leading order in 
		$Q_0=k_{10}-k_{20}$, the CTP counterpart of (\ref{3loopCME}) 
		\begin{align}
			\Lambda_{\rho\lambda}(q)&=-Q_0\frac{e^6}{2\pi^2}\epsilon_{mnl}\sum_{abcd}(-)^{c-1}\int\frac{d^4k}{(2\pi)^4}k_lD^{ac}(k)D^{bc}(k)
			\Gamma_{mn\rho\lambda}^{ab1d}(q,q;k),\nonumber\\
			&=Q_0L_{\rho\lambda}(q).
			\label{3loop_kernel}
		\end{align}
		in terms the CTP photon propagator $D^{ab}(k)\delta_{\mu\nu}$ in the Feynman gauge and the CTP four photon box amplitude $\Gamma_{mnij}^{ab1d}(q,q;k)$ as $Q_0\to 0$. The coefficient 
		of $Q_0$
		\begin{equation}
			L_{\rho\lambda}(q)=-\frac{e^6}{2\pi^2}\epsilon_{mnl}\sum_{abcd}(-)^{c-1}\int\frac{d^4k}{(2\pi)^4}k_lD^{ac}(k)D^{bc}(k)
			\Gamma_{mn\rho\lambda}^{ab1d}(q,q;k),
			\label{3loopkernel}
		\end{equation} 
		with $(\rho, \lambda)=(i, j)$ is the 3-loop contribution to the kernel $K_{ij}(q)$ of the chiral magnetic current current. As is discussed above, 
		$L_{ij}(q)$ in response to a static magnetic field can be evaluated a la Matsubara formulation. We have
		\begin{equation}
			L_{ij}({\bf q})= -\frac{e^6}{2\pi^2}\epsilon_{mnl}T\sum_n\int\frac{d^3{\bf k}}{(2\pi)^3}k_l\frac{1}{(k^2)^2}\Gamma_{mnij}^{(M)}(q,q;k).
			\label{matsubara}
		\end{equation}
		where $k=({\bf k}, -2n\pi T)$ with $n$ an integer, $q=({\bf q},0)$ and the Matsubara amplitude of the 4-photon box diagram reads
		\begin{equation} 
			\Gamma_{mnij}^M(q,q;k)=\rm{I}^{(M)}+\rm{II}^{(M)}+\rm{III}^{(M)}+\rm{IV}^{(M)}+\rm{V}^{(M)}+\rm{VI}^{(M)}.
		\end{equation}
		with $\rm{I}^{(M)}, ... \rm{VI}^{(M)}$ obtained by the following replacement in ${\rm I}, ..., {\rm VI}$ of (\ref{amplitude}),
		\begin{equation}
			p\rightarrow ({\bf p},-(2n+1)\pi T),    \qquad \int\frac{d^4k}{(2\pi)^4}\rightarrow T\sum_n\int\frac{d^3{\bf k}}{(2\pi)^3}.
		\end{equation}
		All derivative formulas at zero temperature can be generalized to the Matsubara amplitude as long as the derivatives is with respect to the spatial components, 
		${\bf q}$ or ${\bf p}$. In particular, we have
		\begin{align}
			\epsilon_{mnl}\Gamma_{nmij}^{(M)}(0,0;k) =& iT\sum\int\frac{d^3{\bf p}}{(2\pi)^3}\frac{\partial}{\partial p_j}\epsilon_{mnl}{\rm tr}\gamma_m
			\left(\frac{1}{p\!\!\!/-{k\!\!\!/}_1}\gamma_n\frac{1}{p\!\!\!/}\gamma_i\frac{1}{p\!\!\!/}
			+\frac{1}{p\!\!\!/}\gamma_i\frac{1}{p\!\!\!/}\gamma_n\frac{1}{p\!\!\!/+k\!\!\!/}\right),\nonumber\\
			=& iT\sum\int\frac{d^3{\bf p}}{(2\pi)^3}\frac{\partial}{\partial p_i}\epsilon_{mnl}{\rm tr}\gamma_m
			\left(\frac{1}{p\!\!\!/-{k\!\!\!/}_1}\gamma_n\frac{1}{p\!\!\!/}\gamma_j\frac{1}{p\!\!\!/}
			+\frac{1}{p\!\!\!/}\gamma_j\frac{1}{p\!\!\!/}\gamma_n\frac{1}{p\!\!\!/+k\!\!\!/}\right).
			\label{leading} 
		\end{align}
		and the coefficient of the linear term in ${\bf q}$ of $\Gamma_{mnij}^{(M)}(q,q,k)$ reads
		\begin{align}
			\left.\frac{\partial}{\partial q_h}\Gamma_{mnij}^{(M)}(q,q;k)\right|_{q=0}
			=&\left.\frac{\partial}{\partial q_{1h}}\Gamma_{mnij}^{(M)}(q_1,0;k)\right|_{q_1=0}+\left.\frac{\partial}{\partial q_{2h}}\Gamma_{mnij}^{(M)}(0,q_2;k)\right|_{q_2=0},\nonumber\\
			=&-T\sum\int\frac{d^3{\bf p}}{(2\pi)^3}\left[\frac{\partial}{\partial p_j}{\rm tr}\gamma_m\left(
			\frac{1}{p\!\!\!/-{k}\!\!\!/}\gamma_n\frac{1}{p\!\!\!/}\gamma_h\frac{1}{p\!\!\!/}\gamma_i\frac{1}{p\!\!\!/}
			-\frac{1}{p\!\!\!/}\gamma_i\frac{1}{p\!\!\!/}\gamma_h\frac{1}{p\!\!\!/}\gamma_n\frac{1}{p\!\!\!/+{k}\!\!\!/}
			\right)-(i\leftrightarrow j)\right].
		\end{align}
	\end{widetext}
	Unlike the zero temperature case, the denominator of a Matsubara Dirac propagator never vanishes because of the nonzero fermionic Matsubara energy. Consequently, 
	the Stokes theorem can be applied safely to the integration of the loop momentum, which yields
	\begin{align}
		\nonumber &\Gamma_{mnij}^{(M)}(0,0;k)=0,\\
		&	\left.\frac{\partial}{\partial q_h}\Gamma_{mnij}^{(M)}(q,q;k)\right|_{q=0}=0.
		\label{partial_q}
	\end{align}
	because the integrand vanishes fast enough as the momentum tends to infinity. Therefore the radiative correction that contributes to the chiral magnetic current at zero temperature no longer contributes when the temperature $T>>q$.

	\subsection{The order $\lim_{\bf Q\to 0}\lim_{Q_0\to 0}$:}

	The first limit $Q_0\to 0$ renders the axial chemical potential static. Given the momentum carried by the external
	vector potential, $q_1=({\bf q}_1,\omega)$ and that carried by the curernt, $q_2=({\bf q}_2,\omega)$, the momentum flowing in through the axial chemical potential is 
	$({\bf q}_2-{\bf q}_1,0)$. Because of the nonzero spatial momentum at the temporal component of the axial vector vertex, the top triangle is no longer tied to 
	the axial anomaly and its dependence on the internal momentum $k$ becomes more complicated than (\ref{anomaly_}). In addition, its contribution 
	may not be diagonal with respect to CTP indices. It follows from the rotation symmetry that the only nonzero component of $\Delta_{\mu\nu4}^{abc}(k,k)$ reads 
	\begin{equation}
		\Delta_{ij4}^{abc}(k,k)=i\frac{e^2}{2\pi^2}\Delta^{abc}({\bf k}^2,k_0)\epsilon_{ijl}k_l,
	\end{equation}
	with $\Delta^{abc}({\bf k}^2,k_0)$ a scalar form factor and we have
	\begin{align}
		\nonumber L_{\rho\lambda}(q)=& -i\frac{e^6}{2\pi^2}\epsilon_{mnl}\sum_{abcd}\int\frac{d^4k}{(2\pi)^4}\Delta^{abc}({\bf k}^2,k_0)\\
		&\times D^{aa'}(k)D^{bb'}(k)\Gamma_{mn\rho\lambda}^{ab1d}(q,q;k).
	\end{align}
	in contrast to (\ref{3loopkernel}) above. Nevertheless, for the standard setup underlying (\ref{classicalCME}), all external momenta are static and the retarded CTP Green's function reduces to the Matsubara one, which 
    shares the same 4-photon box as in the preceding subsection. Following the argument there, in the limit ${\bf q}_2\to {\bf q_1}\equiv {\bf q}$, the term of the 4-photon box that is linear in ${\bf q}$ vanishes because of (\ref{partial_q}) and the three loop diagram in Fig. 1 does not contribute to the chiral magnetic current (\ref{classicalCME}) in this order of limits.

	Though the infrared behavior of the photon or gluon propagator becomes worse in the Matsubara formulation when its energy vanishes, which is the underlying mechanism of the Linde's problem of a pure Yang-Mills gas\cite{Linde},   it does not cause problems here. By turning argument leading to (\ref{partial_q}) around, we can prove that 
	\begin{align}
		\nonumber &\Gamma_{mnij}^{(M)}(q,q;0)=0,\\
		&	\left.\frac{\partial}{\partial k_h}\Gamma_{mnij}^{(M)}(q,q;k)\right|_{k=0}=0.
		\label{partial_k}
	\end{align}
	Combining (\ref{partial_q}) and (\ref{partial_k}), we find that 
	\begin{equation}
		\Gamma_{mnij}^{(M)}=O({\bf k}^2{\bf q}^2).
		\label{IRbox}
	\end{equation}
	at $q_0=k_0=0$ for $|{\bf q}|<<T$ and $|{\bf k}|<<T$. 
	Consequently, the integration of the spatial momentum of the internal photon lines at zero Matsubara energy is infrared safe.
	Applying (\ref{partial_q}) to the four photon box at the bottom of Fig.2, we rule out the radiation corrections to CME from the chain of four photon box in Fig.2 in both orders of the 
	constant axial chemical potential limit.

	While the absence of the radiative correction in the Matsubara formulation follows from the 4D generalization of the Coleman-Hill theorem\cite{ColemanHill} for the 3D QED, the key step here 
	is to recognize that the real time amplitude, where the infrared behavior is difficult to track, becomes Matsubara amplitude after setting all external momenta static and the Coleman-Hill theorem can be applied to the four photon box then because of the benign infrared behavior of the Dirac propagator. The subsequent discussions, after eq.(\ref{matsubara}), provide a concrete implementation of the theorem. See the appendix of cite{ColemanHill} for an alternative proof of (\ref{IRbox}).
	Though the radiative correction brought about by the diagram of Fig.1 does not contribute to the chiral magnetic current in a static and uniform magnetic field, but it does 
	contribute to the current in a static but inhomogeneous magnetic field through higher powers in $|{\bf q}|$. The difference between the two orders of the limit (in susections 3.A and 3.b) as a reminiscence of the real time  Green's function, remains there.

	The kinematic region with the absence of the radiative correction from Fig.1 and 2 to CME, $|{\bf q}|<<T$, does not include the zero temperature point. As the temperature is lowered at a fixed 
	$|{\bf q}|$, the higher powers of $|{\bf q}|$ becomes more important and the radiative corrections to CME are gradually built up until $T<<|{\bf q}|$, then the radiative 
	correction discussed in the preceding section emerges and the difference between the two orders of limits diminishes.

	\section{Summary and Conclusions}

	In this work, we analyzed the impact of a higher order correction to the matrix element of the axial anomaly in massless QED. 
	Through the interplay between the vector Ward identity and the infrared singularity with massless fermions, we are able to re-derive 
	in a much simpler way the known corrections from the three-loop diagram in Fig. 1 at zero temperature. This type of radiative corrections does not invalidate the Adler-Bardeen theorem since the kinematic point required by the theorem cannot be reached 
	for a massless fermion field, but it does contribute to the chiral magnetic current in the massless limit at zero temperature.

	Then we move on to the case of a nonzero temperature where the limit of zero energy-momentum along an external line becomes subtle. 
	We consider two orders of zero energy-momentum limit pertaining to the axial chemical potential: 1) Starting with a homogeneous axial chemical potential and its time dependence is switched off later and 2) Starting with a static axial chemical potential and its inhomogeneity is removed later. Though the real time formulation has to be employed to carry out the zero energy limit, the Matsubara formulation suffices afterward for a static magnetic field, where the infrared singularity of the massless Dirac propagator at zero temperature disappears and the vector Ward identity of the four photon box can be applied without hurdles. Consequently, the three-loop diagram does not contribute a radiative correction to the chiral magnetic current in either order of limit. The same conclusion holds for the higher order diagrams with repeated photon-photon scattering amplitudes in accordance with the Coleman-Hill theorem and the one-loop results (\ref{order1}) and (\ref{order2}) remains intact for the diagrams in Fig.2. The absence of the radiative correction also applies to the three-loop diagram with the two internal photon lines replaced by two gluon lines. Not addressed in this paper is the case with a time dependent magnetic field, which cannot be evaluated with the Matsubara formulation and one has to track the infrared behavior of the full-fleged real time formulation throughout. Denoting the kernel for the response of the electric current to the magnetic field by $K_{ij}^{(s)}(q)$ with $s=1,2$ corresponding the the two orders of limits, 1) and 2), discussed above, our result of CME in massless QCD can be summarized in the following equations
	\begin{equation}
		K_{ij}^{(s)}({\bf q})=i\frac{e^2}{2\pi^2}F_s\left(\frac{|{\bf q}|}{T}\right)\epsilon_{ikj}q_k,
		\label{summary}
	\end{equation} 
	We find that $F_1(|{\bf q}|/T)\to 1$ and $F_2(|{\bf q}|/T)\to 0$ as $|{\bf q}|/T\to 0$ for all QED diagrams in Fig.2 up to $O(g^4)$ QCD correction.	In the low temperature limit, $|{\bf q}|/T\to\infty$,
	\begin{equation}
		F_s(|{\bf q}|/T)\to 1-\frac{3e^4}{64\pi^4}\log\frac{\Lambda^2}{q^2},
	\end{equation}
	for the three-loop QED diagrams of Fig.1 and 
	\begin{equation}
		F_s(|{\bf q}|/T)\to 1-\frac{3g^4}{32\pi^4}\log\frac{\Lambda^2}{q^2}.
	\end{equation}
	if the two internal photon lines in Fig.1 are replaced by gluon lines.

	It is interesting to notice that the difference between the operator anomaly and its matrix element at zero temperature is carried over to the nonzero 
	temperature case. At zero temperature, axial anomaly at the operator level does not require a special kinematic point but the matrix element of the axial 
	anomaly does. So is the case at a nonzero temperature, the chiral magnetic current (\ref{classicalCME}) implied by the one-loop axial anomaly (\ref{anomaly_}), holds for arbitrary space-time dependent magnetic 
	field. Its validity can be extended beyond thermal equilibrium since the identity (\ref{basic}) does not rely on the specific form of the CTP propagator (\ref{ctpS}). 
	With the final state photon-photon scattering, however, the validity of (\ref{classicalCME}) is limited to the low external momenta for massive fermions at zero temperature 
	and is limited to a static and homogeneous magnetic field at a nonzero temperature.
	The temperature here plays a similar role of infrared cutoff as the fermion mass at zero temperature.

	While our analysis supports the absence of radiative correction to the chiral magnetic current under a static and homogeneous magnetic field in massless QED at a nonzero temperature. 
	The QCD corrections, however is much more difficult to assess. 
	Though we are able to rule out the radiative correction from the diagram with the two internal photon lines replaced by two gluon lines, the higher order QCD diagram cannot be 
	decomposed as simple as Fig.2 because of the gluon's self-coupling. Within the framework of Matsubara formulation, the potential infrared singularity comes from the terms with 
	zero Matsubara energies along all internal gluon and photon lines\cite{Linde}. A power counting argument in analoguous to that in\cite{RenHouFeng} yields a contribution of the order
	\begin{equation}
		\left(\frac{|{\bf q}|}{T}\right)^{\frac{1}{2}V_\gamma+\frac{1}{2}V_q-\frac{1}{2}V_3-V_4+1}.
	\end{equation}
	to $F_s(|{\bf q}|/T)$ of (\ref{summary}), where $V_\gamma$, $V_q$, $V_3$ and $V_4$ are the numbers of quark-photon, quark-gluon, 3-gluons and 4-gluons vertices.
	Such an infrared catastrophe with increasing numbers of gluon-self coupling vertices may be regulated by the nonperturbative chromomagnetic mass. A full non-perturbative 
	calculation, say lattice simulation, is yet required to assess the robustness of the classical form of the chiral magnetic current in the presence of the gluon dynamics.

	In non-central heavy ion collisions, the magnetic field created therein is both transient and inhomogeneous. The spatial distribution of the magnetic field is dominated in the plane perpendicular to the reaction plane and the spatial size, according to the simulations\cite{magneticfield1,magneticfield2,magneticfield3}, is about $5fm$ in collisions at center-of-mass energy $\sqrt{s}=200GeV$. Therefore, the spatial momentum carried by the external vector potential $|{\bf q}|$ is roughly about 
	$40MeV$, which is smaller than the temperature $T\sim 200MeV$ reached in the collision. But the life time is about $1\sim 2fm$, making $|{\bf q}|<<q_0$, outside the kinetic region 
	$|{\bf q}|>q_0$ covered here. As discussed in section 3, the real time formulation has to be employed throughout the analysis to tackle the radiative corrections for $|{\bf q}|<<q_0$, 
	and $|{\bf q}|<<T$, in which case the infrared behavior is more convoluted to track. This is currently being explored and the result will be report in near future.

	\begin{acknowledgments}

		We are grateful to D. Kharzeev for bringing  our attention the work \cite{Ansel'm} and the difference between the operator anomaly and its matrix elements. This work is in part supported by the Ministry of Science and Technology of China (MSTC) under the ¡°973¡± Project No. 2015CB856904(4), and by NSFC under Grant Nos. 11735007, 11521064.

	\end{acknowledgments}

	\appendix

	\section{}

The CME current corresponds to the static limit $Q=(q_2-q_1)\rightarrow 0$ of the divergence of the amplitudes of diagrams in Fig. 1. One has
\begin{widetext}
	\begin{align}
		\nonumber    \Lambda_{\rho\lambda}(q_1,q_2)=&-i\frac{e^6}{2\pi^2}\epsilon_{\alpha\mu\beta\nu}\int\frac{d^4p}{(2\pi)^4}\int\frac{d^4k_1}{(2\pi)^4}Q_{\alpha}k_{1\beta}\frac{1}{(k_1^2)^2}\\
		\nonumber & \times {\rm tr}\left[2\frac{p\!\!\!/}{p^2}\gamma_\mu\frac{p\!\!\!/-{k\!\!\!/}_1}{(p-k_1)^2}\gamma_\nu\frac{p\!\!\!/}{p^2}\gamma_\lambda\frac{p\!\!\!/+{q\!\!\!/}_1}{(p+q_1)^2}\gamma_\rho +\frac{p\!\!\!/}{p^2}\gamma_\mu\frac{p\!\!\!/-{k\!\!\!/}_1}{(p-k_1)^2}\gamma_\rho\frac{p\!\!\!/-{k\!\!\!/}_1+{q\!\!\!/}_2}{(p-k_1+q_2)^2}\gamma_\nu\frac{p-q_2}{{p\!\!\!/}-{q\!\!\!/}_2}\gamma_\lambda\right]\\
		&+\left(\rho\leftrightarrow \lambda, q_1\leftrightarrow q_2 \right)+{\cal O}(Q^2).
	\end{align}
\end{widetext}
Here we explicitly present the permutation factor connecting the photon-box to the triangle diagrams. Note that the divergence of AVV triangle digram can be regarded as a point and represented by $e^2/(2\pi^2)\epsilon_{\alpha\mu\beta\nu}Q_{\alpha}k_{1\beta}$, which enables us to ignore completely the $q$ dependence in the photon-box and the internal two photon lines in the leading order of $Q$.  In the following calculations, we only aim to keep the UV divergent part and will ignore the finite terms entirely.  For this purpose, it is in great advantage to carry out firstly the integral of the photon-box loop momentum $p$  and then that of the internal photon momentum $k_1$, because the former integral of the momentum $p$ turns out to be finite.  

It can be shown by simple power counting that the potentially UV divergent part in integral of $p$  reads
\begin{equation}
\epsilon_{\alpha\sigma\beta\rho}\int\frac{d^4p}{(2\pi)^4}\frac{1}{(p^2)^4}{\rm tr}\left[\gamma_\rho p\!\!\!/\gamma_\delta p\!\!\!/\gamma_\sigma p\!\!\!/\gamma_\lambda p\!\!\!/+\gamma_\rho p\!\!\!/\gamma_\lambda p\!\!\!/\gamma_\delta p\!\!\!/\gamma_\sigma p\!\!\!/\right],\label{pintegral}
\end{equation} 
Symmetry allows one to replace
\begin{equation}
p_\alpha p_\beta p_\mu p_\nu\sim \frac{1}{24}(p^2)^2(\delta_{\alpha\beta}\delta_{\mu\nu}+\delta_{\alpha\mu}\delta_{\beta\nu}+\delta_{\alpha\nu}\delta_{\beta\mu}),
\end{equation}
The vanishing of the first term  in eq. (\ref{pintegral}), corresponding to the photon-box diagram (b) in Fig.1, can be shown by noting that
\begin{align}
	\nonumber (\delta_{\alpha\beta}\delta_{\mu\nu}+\delta_{\alpha\mu}\delta_{\beta\nu}&+\delta_{\alpha\nu}\delta_{\beta\mu})\epsilon_{\xi\sigma\kappa\rho}\\
	&\times{\rm tr}(\gamma_\rho\gamma_\alpha\gamma_\delta\gamma_\beta\gamma_\sigma\gamma_\mu\gamma_\lambda\gamma_\nu)=0,
\end{align}
Likewisely, the same photon-box diagram $(b)$ in Fig. 1 with two final photon states interchanged is also finite. The similar factor for the second term in eq.(\ref{pintegral}) is, however, nonzero
\begin{align}
	\nonumber (\delta_{\alpha\beta}\delta_{\mu\nu}&+\delta_{\alpha\mu}\delta_{\beta\nu}+\delta_{\alpha\nu}\delta_{\beta\mu})\epsilon_{\xi\sigma\kappa \rho}\\
	&\times{\rm tr}(\gamma_\rho\gamma_\alpha\gamma_\lambda\gamma_\beta\gamma_\delta\gamma_\mu\gamma_\sigma\gamma_\nu)=96\epsilon_{\xi\delta\kappa\lambda}.\label{figurea}
\end{align}
corresponding to the photon-box diagram $(a)$ in Fig. 1. But the sum of (\ref{figurea}) and the one with two final photon states interchanged is manifestly finite due to Bose symmetry. 

Performing the finite integral of momentum $p$ and keeping only the UV divergent terms upon the integral of $k_1$, one has
\begin{widetext}
	\begin{align}
		\nonumber  & \Lambda_{\rho\lambda}(q_1,q_2)=\frac{e^6}{8\pi^4}Q_{\alpha}q_{1\beta}\epsilon_{\alpha\rho \beta\lambda} \int^\Lambda\frac{d^4k_1}{(2\pi)^4}\frac{1}{(k_1^2)^2}\int_0^1 dy\int_0^{1-y}dw(1-y-w)\\
		& \ \ \ \ \ \ \ \ \ \ \ \ \ \ \  \times\left[\frac{12yw^2-22yw-6w^2+9w+9y-3}{w(w-1)^2}+\frac{1}{(y+w)(y+w-1)}\right]+\left(\lambda\leftrightarrow \rho, q_1\leftrightarrow q_2 \right).
	\end{align}
\end{widetext}
with $\Lambda$ an UV cutoff. We note that the result is logarithmically divergent. Carrying out the integral of Feynman parameters, one obtains
\begin{align}
	\nonumber   \Lambda_{\rho\lambda}(q_1,q_2)=&\frac{-3 e^6}{8\pi^4}Q_{\alpha}q_{1\beta}\epsilon_{\alpha\rho \beta\lambda} \int^\Lambda\frac{d^4k_1}{(2\pi)^4}\frac{1}{(k_1^2)^2},\\
	=& i\frac{3 e^6}{128\pi^6}Q_{\alpha}q_{1\beta}\epsilon_{\alpha\rho \beta\lambda}\ln\frac{\Lambda^2}{k^2}.
\end{align}	
with $k$ an infrared cutoff depending on the external momenta.


	\section{}

	In this appendix, we discuss the relation between the real time and Matsubara formulations of a three point Green's function. In terms of the 
	spectral representation, our result applies to all orders of perturbation theory.  Let $A(t_1)$, $B(t_2)$
	and $C(t_3)$ be field operators at real times $t_1$, $t_2$ and $t_3$ in the Heisenberg representation with the time evolution
	\begin{equation}
		O(t)=e^{iHt}O(0)e^{-iHt}
	\end{equation}
	where $H$ is the Hamiltonian of the system. The different symbols, $A$, $B$ and $C$ reflect different operators or the
	same operators evaluated at different spatial coordinates. For CME discussed in this work, $A$ and $B$ stand for the electric current 
	density with $B$ attached to the external electromagnetic potential and $C$ represents the axial current density attached to the external 
	axial vector potential. It follows from the CTP formulation that the electric current measured at time $t_1$ in response to the external field at $t_2$ and $t_3$ is given by the three point retarded Green's function
	\begin{equation}
		G_R(t,t')\equiv G_{111}(t,t')-G_{121}(t,t')-G_{112}(t,t')+G_{122}(t,t')
	\end{equation}
	where $t \equiv t_1-t_2$, $t' \equiv t_1-t_3$ and the different CTP components on RHS are given explicitly by
	\begin{eqnarray}
		G_{111}(t,t') &=& <T(A_1B_2C_3)>\nonumber\\
		G_{121}(t,t') &=& <B_2T(A_1C_3)>\nonumber\\
		G_{112}(t,t') &=& <C_3T(A_1B_2)>\nonumber\\
		G_{122}(t,t') &=& <\tilde T(B_2C_3)A_1>
	\end{eqnarray}
	with $A_1\equiv A(t_1)$, $B_2\equiv B(t_2)$, $C_3\equiv C(t_3)$ and $T$ ($\tilde T$) imposes time (anti-time) ordering. The expectation value
	\begin{equation}
		<O>\equiv\frac{{\rm Tr}e^{-\beta H}O}{{\rm Tr}e^{-\beta H}}
	\end{equation}  
	with $\beta$ the inverse temperature and $Z\equiv {\rm Tr}e^{-\beta H}$. 
	Expanding the time and anti-time ordering product, we have
	\begin{widetext}
		\begin{align}
			\nonumber G_R(t,t') =& <A_1B_2C_3>\theta_{123}+<B_2C_3A_1>\theta_{231}+<C_3A_1B_2>\theta_{312}
			+<B_2A_1C_3>\theta_{213}+<A_1C_3B_2>\theta_{132}\\
			\nonumber &+<C_3B_2A_1>\theta_{321}
			- <B_2A_1C_3>\theta_{13}-<B_2C_3A_1>\theta_{31}-<C_3A_1B_2>\theta_{12}
			-<C_3B_2A_1>\theta_{21}\\ 
			&+<C_3B_2A_1>\theta_{23}+<B_2C_3A_1>\theta_{32}
			\label{expansion}
		\end{align}
		where $\theta_{ab}\equiv\theta(t_a-t_b)$ and $\theta_{abc}\equiv\theta(t_a-t_b)\theta(t_b-t_c)=\theta_{ab}\theta_{bc}$
		The spectral representation amounts to inserting the complete set of eigenstates of $H$, $H|N>=E_N|N>$, with
		\begin{equation}
			<N|M>=\delta_{NM} \qquad  1=\sum_N|N><N|
		\end{equation}
		for instance
		\begin{eqnarray}
			<A_1B_2C_3> &=& \frac{1}{Z}\sum_{N,M,L}e^{-\beta E_N}<N|A(t_1)|M><M|B(t_2)|L><L|C(t_3)|N>\nonumber\\
			&=& \frac{1}{Z}\sum_{N,M,L}e^{-\beta E_N}<N|A(0)|M><M|B(0)|L><L|C(0)|N>e^{i(E_N-E_M)t_1+i(E_M-E_L)t_2+i(E_L-E_N)t_3}\nonumber\\
			&=& \frac{1}{Z}\sum_{N,M,L}e^{-\beta E_N}(ABC)e^{i(E_N-E_L)t'+i(E_L-E_M)t}
			\label{spectral}
		\end{eqnarray} 
		Applying this procedure to all terms of (\ref{expansion}) and then transforming to the energy representation, we obtain that
		\begin{align}
			{\cal G}_R(\omega,\omega') \equiv &\int_{-\infty}^\infty dt\int_{-\infty}^\infty dt'e^{i(\omega t+\omega't')}G_R(t,t')\nonumber\\
			= &\frac{1}{Z}\sum_{N,M,L}(ABC)\left[\frac{e^{-\beta E_N}}{(\omega'+E_N-E_L+i0^+)(\omega+\omega'+E_N-E_M+i0^+)}\right.\nonumber\\
			&+ \left.\frac{e^{-\beta E_M}}{(\omega+E_L-E_M+i0^+)(\omega+\omega'+E_N-E_M+i0^+)}
			-\frac{e^{-\beta E_L}}{(\omega'+E_N-E_L+i0^+)(\omega+E_L-E_M+i0^+)}\right]\nonumber\\
			&- \frac{1}{Z}\sum_{N,M,L}(CBA)\left[\frac{e^{-\beta E_N}}{(\omega'+E_M-E_N+i0^+)(\omega+\omega'+E_L-E_N+i0^+)}\right.\nonumber\\
			&+\left.\frac{e^{-\beta E_L}}{(\omega+E_L-E_M+i0^+)(\omega+\omega'+E_L-E_N+i0^+)}
			-\frac{e^{-\beta E_M}}{(\omega'+E_M-E_N+i0^+)(\omega+E_L-E_M+i0^+)}\right]
		\end{align}
		after some cyclic permutations under the trace, where
		\begin{eqnarray}
			(ABC) &\equiv& <N|A(0)|M><M|B(0)|L><L|C(0)|N>\nonumber\\
			(CBA) &\equiv& <N|C(0)|M><M|B(0)|L><L|A(0)|N>.
		\end{eqnarray}
		The poles on the complex-$\omega$ or complex-$\omega'$ plane are all located below the real axis as expected. In the static limit, we find that
		\begin{align}
			{\cal G}_R(0,0) =& -\frac{1}{Z}\sum_{N,M,L}[(ABC)+(BCA)]\left[\frac{e^{-\beta E_N}}{(E_N-E_L)(E_N-E_M)}\right.\nonumber\\
			&+\left.\frac{e^{-\beta E_M}}{(E_M-E_L)(E_M-E_N)}+\frac{e^{-\beta E_L}}{(E_L-E_M)(E_L-E_N)}\right]
			\label{static}
		\end{align}
		There are no sigularities when any pair of $E_N$, $E_M$ and $E_L$ coalesce.

		Coming to the Matsubara formulation, the Euclidean time evolution of an operator is generated by
		\begin{equation}
			O(\tau)=e^{H\tau}O(0)e^{-H\tau}
		\end{equation}
		with $0\le\tau<\beta$. The three point Green's function corresponding to (\ref{expansion}) reads
		\begin{align}
			G_M(\tau_1,\tau_2\tau_3) =& <{\cal T}[A(\tau_1)B(\tau_2)C(\tau_3)]>\nonumber\\
			=& \frac{1}{Z}[{\rm Tr}e^{-\beta H}A(\tau_1)B(\tau_2)C(\tau_3)\theta(\tau_1-\tau_2)\theta(\tau_2-\tau_3)+ \hbox{5 other permutations}]
		\end{align}
		and its Fourier transformation at Matsubara energies $\omega_1$, $\omega_2$ and $\omega_3$
		\begin{equation} 
			{\cal G}_M(\omega_1,\omega_2,\omega_3)=\int_0^\beta d\tau_1\int_0^\beta d\tau_2\int_0^\beta d\tau_3 
			e^{i(\omega_1\tau_1+\omega_2\tau_2+\omega_3\tau_3)}G_M(\tau_1,\tau_2\tau_3)
		\end{equation}
		In particular, at zero external Matsubara energies
		\begin{equation}
			{\cal G}_M(0,0,0)=\int_0^\beta d\tau_1\int_0^\beta d\tau_2\int_0^\beta d\tau_3 G_M(\tau_1,\tau_2\tau_3).
			\label{static_M}
		\end{equation}
		Inserting the complete set of eigenstates of $H$ as (\ref{spectral}) and carrying out the integral (\ref{static_M}), we end up with
		\begin{align}
			{\cal G}_M(0,0,0) =& \frac{\beta}{Z}\sum_{N,M,L}[(ABC)+(BCA)]\left[\frac{e^{-\beta E_N}}{(E_N-E_L)(E_N-E_M)}\right.\nonumber\\
			&+\left.\frac{e^{-\beta E_M}}{(E_M-E_L)(E_M-E_N)}+\frac{e^{-\beta E_L}}{(E_L-E_M)(E_L-E_N)}\right].
			\label{static}
		\end{align}
		Comparing with (\ref{static}), we find that
		\begin{equation}
			{\cal G}_M(0,0,0)=-\beta{\cal G}_R(0,0).
		\end{equation}		
	\end{widetext}

	\section{}

	Using the identity
	\begin{equation}
		\frac{1}{p\!\!\!/+q\!\!\!/}q\!\!\!/\frac{1}{p\!\!\!/}=\frac{1}{p\!\!\!/}-\frac{1}{p\!\!\!/+q\!\!\!/}.
	\end{equation}	
	we find that
	\begin{align}
		\nonumber -iq_{2\lambda}\Gamma_{\mu\nu\rho\lambda}(q_1,q_2;k_1)=&\int\frac{d^4p}{(2\pi)^4}\Big[G_{\mu\nu\rho}(p+q_2|q_1,q_2;k_1)\\
		&-G_{\mu\nu\rho}(p|q_1,q_2;k_1)\Big],	
	\end{align}
	where
	\begin{align}
		\nonumber G_{\mu\nu\rho}(p|q_1,q_2;k_1)=&{\rm tr}\gamma_\mu\left(\frac{1}{p\!\!\!/-{k\!\!\!/}_1}\gamma_\nu\frac{1}{p\!\!\!/-{q\!\!\!/}_2+{q\!\!\!/}_1}
		\gamma_\rho\frac{1}{p\!\!\!/-{q\!\!\!/}_2}\right.\\
		&+\left.\frac{1}{p\!\!\!/}\gamma_\rho\frac{1}{p\!\!\!/-{q\!\!\!/}_1}\gamma_\nu\frac{1}{p\!\!\!/-{q\!\!\!/}_2+{k\!\!\!/}_1}\right).
	\end{align}
	and
	\begin{align}
		\nonumber -iq_{1\rho}\Gamma_{\mu\nu\rho\lambda}(q_1,q_2;k_1)=&\int\frac{d^4p}{(2\pi)^4}\Big[H_{\mu\nu\lambda}(p+k_1|q_1,q_2;k_1)\\
		\nonumber &-H_{\mu\nu\lambda}(p|q_1,q_2;k_1)\\
		\nonumber &+I_{\mu\nu\lambda}(p-k_1+q_1|q_1,q_2;k_1)\\
		&-I_{\mu\nu\lambda}(p|q_1,q_2;k_1)\Big],
	\end{align}
	where
	\begin{align}
		\nonumber H_{\mu\nu\lambda}(p|q_1,q_2;k_1)=&{\rm tr}\gamma_\mu\left(\frac{1}{p\!\!\!/-{k\!\!\!/}_1}\gamma_\nu\frac{1}{p\!\!\!/+{q\!\!\!/}_2}\gamma_\lambda\frac{1}{p\!\!\!/}\right)\\
		&+{\rm tr}\gamma_\mu\left(\frac{1}{p\!\!\!/-{k\!\!\!/}_1}\gamma_\lambda\frac{1}{p\!\!\!/-{k\!\!\!/}_1+{q\!\!\!/}_2}\gamma_\nu\frac{1}{p\!\!\!/}\right),
	\end{align}
	and
	\begin{align}
		\nonumber I_{\mu\nu\lambda}(p|q_1,q_2;k_1)=&{\rm tr}\gamma_\mu\left(\frac{1}{p\!\!\!/-{q\!\!\!/}_1}\gamma_\nu\frac{1}{p\!\!\!/+{k\!\!\!/}_1-{q\!\!\!/}_2}\gamma_\lambda\frac{1}{p\!\!\!/+{k\!\!\!/}_1}
		\right)\\
		&+{\rm tr}\gamma_\mu\left(\frac{1}{p\!\!\!/-{q\!\!\!/}_1}\gamma_\lambda\frac{1}{p\!\!\!/-{q\!\!\!/}_1-{q\!\!\!/}_2}\gamma_\nu\frac{1}{p\!\!\!/+{k\!\!\!/}_1}
		\right).
	\end{align}
	Both $G_{\mu\nu\rho}(p|q_1,q_2;k_1)$ and $H_{\mu\nu\lambda}(p|q_1,q_2;k_1)$ give rise to linearly divergent integrals by power counting and a regulator is required to justify the shift of the integration momentum $p$ and thereby the Ward identity. On the other hand, the leading terms in $p$ are canceled in the combinations
	$\epsilon_{\mu\nu\alpha\lambda}G_{\mu\nu\rho}(p|q_1,q_2;k_1)$ and $\epsilon_{\mu\nu\alpha\beta}H_{\mu\nu\rho}(p|q_1,q_2;k_1)$ and the integration becomes logarithmic 
	divergence by power counting. Consequently, the Ward identities
	\begin{align}
		&-iq_{2\lambda}\epsilon_{\mu\nu\alpha\beta}\Gamma_{\mu\nu\rho\lambda}(q_1,q_2;k_1)= 0,\\
		&-iq_{1\rho}\epsilon_{\mu\nu\alpha\beta}\Gamma_{\mu\nu\rho\lambda}(q_1,q_2;k_1)=0.
	\end{align}
	without the need of regulators.

	\section{}

		The identities underlying the vector and axial vector (anomalous) Ward identity
	\begin{align}
		S_F(p+q)q\!\!\!/S_F(p)= & i[S_F(p)-S_F(p+q)],\nonumber\\
		\nonumber	S_F(p+q)q\!\!\!/\gamma_5S_F(p)=& i[\gamma_5S_F(p)+S_F(p+q)\gamma_5]\\
		&+2mS_F(p+q)\gamma_5S_F(p).\label{identity}
	\end{align}
	with $S_F(p)$ the free Dirac propagator in Feynman diagrams can be readily generalized to the CTP formulation.
	Consider the integral
	\begin{equation}
		I_V\equiv \int_{\cal C}d^4y\frac{\partial\theta}{\partial y_\mu}S(x-y)\gamma_\mu S(y-x'),
	\end{equation}
	where
	$S(x-y)\equiv \langle\psi(x)\bar\psi(y)\rangle$ with $\langle...\rangle$ the average with respect to the CTP path integral with aribitrary density operator and $\theta(x)$ is an arbitrary 
	function. The time components of all coordinates are along the closed path of the CTP formulation with 
	\begin{equation}
		\int_{\cal C}d^4y(...)\equiv\int d^3{\bf y}\left(\int_{-\infty}^\infty+\int_\infty^{-\infty}\right)dt(...).
	\end{equation}
	By definition
	\begin{equation}
		-\left(\gamma_\mu\frac{\partial}{\partial x_\mu}+m\right)S(x-y)=i\delta^4(x-y),
	\end{equation}
	and
	\begin{equation}
		S(x-y)\left(\overleftarrow{\gamma_\mu\frac{\partial}{\partial y_\mu}}-m\right)=i\delta^4(x-y).
	\end{equation}

	Integrating by part, we find that
	\begin{equation}
		I_V=-i[\theta(x)-\theta(x')]S(x-x'),
		\label{vector}
	\end{equation}
	Labeling different branches of the closed time path by different CTP indices, $S(x-y)$ becomes a $2\times 2$ matrix with respect to CTP indices and (\ref{vector}) becomes
	\begin{align}
		\nonumber \int_{\cal C}d^4y\frac{\partial\theta_c}{\partial y_\mu}S_{ac}(x-y)\gamma_\mu S_{cb}(y-x')=&\theta_c(x)S_{cb}(x-x')\\
		&-S_{ac}(x-x')\theta_c(x'),
	\end{align}
	or
	\begin{align}
		\nonumber \int_{\cal C}d^4y\frac{\partial\theta}{\partial y_\mu}S(x-y)\eta_c\gamma_\mu S(y-x')
	=&\theta(x)\Big[\eta_cS(x-x')\\
		&-S(x-x')\eta_c\Big].
	\end{align}
	with $\eta_c$ the projection operator defind in (\ref{projection}). Transforming to the momentum space, we end up with the generalization of the first identity of (\ref{identity})
	\begin{equation}
		S(p+q)Q{\eta\!\!\!/}_cS_F(p)=i[\eta_cS(p)-S(p+q)\eta_c],
	\end{equation}
	Applying the same technique to the integral
	\begin{equation}
		I_A\equiv \int_{\cal C}d^4y\frac{\partial\theta}{\partial y_\mu}S(x-y)\gamma_\mu\gamma_5 S(y-x').
	\end{equation}
	we obtain eq.(\ref{basic}) as the generalization of the second identity of (\ref{identity}).

	\newpage 


\end{document}